\documentclass[twocolumn,pra,showpacs,preprintnumbers,amsmath,amssymb]{revtex4}

\usepackage{graphicx}
\usepackage{dcolumn}
\usepackage{bm}
\usepackage{booktabs}

\makeatletter
\AtBeginDocument{\@ifpackageloaded{natbib}{\ifNAT@numbers\if@filesw\immediate\write\@auxout{\string\global\string\NAT@numberstrue}\fi\fi}{}}
\makeatother
\begin{document}

\preprint{APS/123-QED}

\title{Counter-rotating vortices in miscible two-component Bose-Einstein condensates}

\author{Shungo Ishino $^1$, Makoto Tsubota $^{1,2}$, and  Hiromitsu Takeuchi $^1$}
\affiliation{$^1$Department of Physics, Osaka City University, 3-3-138 Sugimoto, Sumiyoshi-ku, Osaka 558-8585, Japan   \\
$^2$ The OCU Advanced Research Institute for Natural Science and Technology (OCARINA),Osaka City University, 3-3-138 Sugimoto, Sumiyoshi-ku, Osaka 558-8585, Japan}

\date{\today}

\begin{abstract}
  Counter-rotating vortices in miscible two-component Bose-Einstein condensates, in which superflows counter-rotate between the two components around the overlapped vortex cores, are studied theoretically in a pancake-shaped potential.
  In a linear stability analysis with the Bogoliubov--de Gennes model, we show that counter-rotating vortices are dynamically unstable against splitting into multiple vortices.
  The instability shows characteristic behaviors as a result of countersuperflow instability, which causes relaxation of relative flows between the two components in binary condensates.
  The characteristic behaviors are completely different from those of multiquantum vortices in single-component Bose-Einstein condensates; the number of vortices generated by the instability can become larger than the initial winding number of the counter-rotating vortex.
  We also investigate the nonlinear dynamics of the instability by numerically solving the Gross-Pitaevskii equations.
  The nonlinear dynamics drastically changes when the winding number of counter-rotating vortices becomes larger, which lead to nucleation of vortex pairs outside of the vortex core.
  The instability eventually develops into turbulence after the relaxation of the relative rotation between the two components.
\end{abstract}

\pacs{
03.75.Mn, 
67.85.De, 
67.25.dk  
}

\maketitle

\section{Introduction}
  Quantized vortices are one of the remarkable consequences of Bose-Einstein condensation and superfluidity in quantum fluids and are  found in superfluids $^4$He and $^3$He and Bose-Einstein condensates (BECs) of atomic gas.
  In the context of hydrodynamics, quantized vortices often appear and play an important role in the understanding of various phenomena, such as the rotating of superfluid He \cite{Donnelly_Book1991}, thermal counterflow in superfluid $^4$He \cite{Vinen1957,Adachi_PRB2010}, and quantum turbulence \cite{Adachi_PRB2010,Kobayashi_PRL2005,Kobayashi_PRA2007}.

  Among the many types of physics of quantized vortices, multiquantum vortices, whose winding number is larger than unity, are an interesting and important subject.
  Multiquantum vortices have never been realized in superfluid $^4$He except in some transient states \cite{Karn_PRB1980}.
  This is chiefly because an $L$-charged vortex, whose winding number is $L$, is unstable and easily splits into $L$ single-quantum vortices, whose winding number is unity.

  Atomic BECs form another subject in the study of multiquantum vortices.
  In experiments,  optical technology enables us to make a multiquantum vortex and visualize the vortex directly \cite{Matthews_PRL1999,Shibayama_JPB2011}.
  Furthermore, because of the weak interaction between the atoms, it is relatively easy to perform a theoretical analysis by using the Gross-Pitaevskii (GP) model and the Bogoliubov--de Gennes (BdG) model \cite{Pethick_book}.
  Thus, the splitting of multiquantum vortices has been experimentally observed \cite{Isoshima_PRL2007,Shin_PRL2004} and theoretically studied \cite{Mottonen_PRA2003,Kawaguchi_PRA2004,Huhtamaki_PRL2006,Isoshima_PRL2007}.
  An $L$-charged vortex essentially has unstable modes with $l(\leqq L)$-fold symmetry and splits into $L$ single-quantum vortices.
  Some studies also discuss multiquantum vortices in two-component BECs \cite{Skryabin_PRA2000,Park_PRAR2004,Brtka_PRA2010,Wen_PRA2013}.
  Such vortex splitting instabilities are understood as a dynamic instability in the BdG model \cite{Pethick_book}.

   Hydrodynamic instability has been actively studied, independent of these topics, in two-component BECs, for example, the Kelvin-Helmholtz instability \cite{Takeuchi_PRB2010,Suzuki_PRA2010} and the Rayleigh-Taylor instability \cite{Sasaki_PRA2009,Gautam_PRA2010}.
  As another type of hydrodynamic instability, we previously studied instability in uniform countersuperflow, miscible two-component BECs with relative superfluid velocity between the two components \cite{Takeuchi_PRL2010,Ishino_PRA2011}.
  It is well known that uniform, miscible two-component BECs are stable when the intraspecies interaction coefficients $g_{11}$ and $g_{22}$ and interspecies interaction coefficient $g_{12}$ satisfy  the condition $g_{11}g_{22}>g_{12}^2$ \cite{Pethick_book}.
  However, when the relative superfluid velocity between the two components exceeds a critical value, the system becomes dynamically unstable, causing a characteristic density pattern and vortex nucleation \cite{Takeuchi_PRL2010, Ishino_PRA2011}.
  The nucleated vortices are stretched so as to reduce the relative superflows between the two components.
  Then, reconnection frequently occurs between the vortices, leading to binary quantum turbulence.
  CSI has been recently observed in experiments \cite{Hamner_PRL2011}.

  In this paper, we discuss counter-rotating (CR) vortices in miscible two-component BECs trapped by a harmonic oscillator potential.
  We consider that the first and second components simultaneously have an $L$-charged vortex and a $-L$-charged one at the center of the BECs, respectively.
  The winding numbers of the two vortices have the same magnitude but  opposite sign.
  Therefore, the two BECs relatively rotate.
  For the following discussion, we denote an $L$-charged vortex in the first and second components of the two-component BEC as $(L,0)$-vortex and $(0,L)$-vortex, respectively.
  Thus, a CR vortex that is overlapped by an $L$-charged vortex and a $-L$-charged vortex is written as an $(L,-L)$-vortex.
  The BECs with a CR vortex are expected to be closely related to countersuperflow because the  BECs with a CR vortex and countersuperflow have  similarity, such as  relative motion.
  Counter-rotating binary BECs have been theoretically studied in a toroidal trap \cite{Suzuki_PRA2010,Abad_arXiv2013}.
  Our work focuses on natures of counter-rotating systems as a vortex in a harmonic oscillator potential.

  This paper is organized as follows.
  In Sec. II, we formulate a system of two-component BECs with a CR vortex in the GP model at zero temperature.
  Section III is devoted to a linear stability analysis of CR vortices in the BdG model.
  We show that the instability of CR vortices is characterized by countersuperflow instability (CSI) by numerically solving the BdG equations.
  In Sec. IV, we reveal the nonlinear development of the instability of CR vortices by numerically solving the time-dependent GP equations.
  The results are summarized in Sec. V.

\section{Formulation}\label{sec:formulations}
  We consider miscible two-component BECs described by the condensate wave functions $\Psi _j({\bm r},t)=\sqrt{n_j({\bm r},t)}e^{i\phi _j({\bm r},t)}$ in the mean-field approximation at zero temperature, where the index $j$ refers to each component ($j=1,2$).
  The wave functions are governed by the coupled GP equations \cite{Pethick_book}
  \begin{eqnarray}
    i \hbar \frac{\partial}{\partial t} \Psi _j = \left(-\frac{\hbar^2}{2m_j}{\bm \nabla}^2+V_j({\bm r})+\sum_{k=1,2} g_{jk}|\Psi _k|^2\right)\Psi _j,
    \label{eq:GP}
  \end{eqnarray}
  where $m_j$ is the mass of the $j$th component and the coefficient $g_{jk}=2\pi\hbar^2a_{jk}/m_{jk}$ represents the atomic interaction with $m_{jk}^{-1}=m_{j}^{-1}+m_{k}^{-1}$ and the $s$-wave scattering length $a_{jk}$ between the $j$th and $k$th components.
  Our analysis supposes the conditions $g_{11}g_{22}>g_{12}^2$ and $g_{jj}>0$, indicating that the static, miscible two-component BECs are stable \cite{Pethick_book}.
  For simplicity, we set the mass and the $s$-wave scattering length of the two components to the same value, namely, $m_1=m_2=m$, $a_{11}=a_{22}=a$, and $g_{11}=g_{22}=g$.
  The particle numbers of the two components are also the same: $N_1=N_2=N$.
  The external trapping potential is a harmonic oscillator potential, given by $V_j({\bm r})=\frac{1}{2}m(\omega_r^2r^2+\omega_z^2z^2)$ with $r^2=x^2+y^2$.

  The BECs may be treated as a two-dimensional system when we use the ``pancake" trap geometry with $\omega_r \ll \omega_z$.
  Therefore, we separate the degrees of freedom of the wave functions as $\Psi_j(x,y,z,t)=\psi_j(x,y,t)\phi_j(z)$.
  When the potential energy in the $z$ direction is sufficiently larger than the interaction energy, $\phi_j(z)$ is approximated by the one-particle ground-state wave function in a harmonic oscillator potential: $\phi_j(z)=\left(N_z/\sqrt{\pi}a_{hz}\right)^{1/2}\exp\left(-{z^2}/{2a_{hz}^2}\right)$, where $N_z$ is a normalization constant and $a_{hz}=\sqrt{\hbar/m\omega_z}$.
  Then the GP equations are reduced to the dimensionless form
  \begin{eqnarray}
    i \frac{\partial}{\partial t} \psi _j = \left(-\frac{1}{2}{\bm \nabla}_r^2+\frac{1}{2}r^2+\sum_{k=1,2} C_{jk}|\psi _k|^2\right)\psi _j
    \label{eq:2DGP}
  ,\end{eqnarray}
  with ${\bm \nabla}_r^2=\partial_r^2+r^{-1}\partial_r-r^{-2}\partial_\theta^2$, where the length, time, and wave functions are scaled as
  \begin{eqnarray}
    x=a_{hr}\tilde{x},\ t=\frac{\tilde{t}}{\omega_r},\ \psi_j=\frac{\sqrt{N_{2D}}}{a_{hr}}\tilde{\psi}_j.
    \label{eq:scale}
  \end{eqnarray}
  Here, $a_{hr}=\sqrt{\hbar/m\omega_r}$ and the two-dimensional particle number $N_{2D}$ relates to $N_z$ through $N=N_{2D}N_z$.
  The tildes in Eq. (\ref{eq:2DGP}) are omitted for simplicity.
  The nondimensional interaction coefficient
  \begin{eqnarray}
    C_{jk}=\frac{2\sqrt{2\pi}Na_{jk}}{a_{hz}}
    \label{eq:Cjk}
  \end{eqnarray}
  includes all parameters of this system.
  Because  the parameters of the two components are the same, the intraspecies interaction coefficients are the same: $C_{11}=C_{22}=C$.
  The chemical potential $\mu_{\rm b}$ and the healing length $\xi_{\rm b}$ of this system are calculated in the Thomas-Fermi approximation \cite{Pethick_book} as
  \begin{align}
    \mu_{\rm b} &= \frac{15^{2/5}}{2}\left(\frac{N(a+a_{12})}{{\bar a}_h}\right)^{2/5}\hbar{\bar \omega}, \\
    \xi_{\rm b}&=\frac{\hbar}{\sqrt{m(g+g_{12})n_{\rm b}}},
  \end{align}
  where ${\bar a}_h\equiv(a_{hr}^2a_{hz})^{1/3}$, ${\bar \omega}\equiv(\omega_r^2\omega_z)^{1/3}$, and $n_{\rm b}$ is the density in bulk.

  The stationary state of two BECs that have an $(L,-L)$-vortex at the center is described by the cylindrically symmetric functions
  \begin{eqnarray}
    \psi_1^0&=\sqrt{n^0_{1}(r)}e^{i\left( L\theta-\mu_{1}t/\hbar\right)}, \label{eq:stationary1} \\
    \psi_2^0&=\sqrt{n^0_{2}(r)}e^{i\left(-L\theta-\mu_{2}t/\hbar\right)}, \label{eq:stationary2}
  \end{eqnarray}
  where $\theta$ is the polar angle and $\mu_j$ is the chemical potential of the $j$th component.
  The square of the amplitude $n^0_{j}(r)$ gives the radial density profile.
  The amplitudes are obtained through the imaginary time propagation method by inserting  Eqs. (\ref{eq:stationary1}) and (\ref{eq:stationary2}) into the GP equations.
  The densities of the two BECs must vanish at $r=0$ and $r=\infty$ because of the vortices and the trapping potential.
  Note that $\mu_1=\mu_2=\mu$ and the densities have the same function, $n^0_{1}(r)=n^0_{2}(r)=n^0(r)$, because the two BECs have the same parameters and the same winding number magnitude.

  In this paper, the nondimensional interaction parameters are $C\simeq 5500$ and $C_{12}=0.9C$, which causes repulsive interspecies interaction.
  For example, the parameters are realized in a system with $10^5$ atoms of $^{87}$Rb for each component.
  The $s$-wave scattering lengths are $a=5.3$  nm and $a_{12}=0.9a$, and also the trapping frequencies are $\omega_r=2\pi\times 5$ Hz and $\omega_z=2\pi\times 500$  Hz.
  Then the healing length is $\xi_{\rm b} = 0.54\; {\mu\rm m}$, where the density $n_{\rm b}$ in bulk is estimated by the Thomas-Fermi approximation without vortices.
  We investigate the cases of small and large winding numbers of CR vortices.

\section{Linear Stability} \label{sec:linear}

  Here, we study the linear stability of the CR vortices in the BdG model.
  After the formulation of the BdG equations, we first discuss the dynamic instability of $(L,-L)$-vortices for small $L$.
  Then we investigate the instability for large $L$ and its relation to CSI.

\subsection{Bogoliubov--de Gennes analysis} \label{sec:bdg}

  We consider a collective excitation above the stationary state written by Eqs. (\ref{eq:stationary1}) and (\ref{eq:stationary2}) as $\psi_j=\psi_j^0+\delta\psi_j$.
  Because the system has  rotational symmetry, we write the excitation wave functions $\delta\psi _{j}$ with
  \begin{align}
    \delta\psi_1&=\ \ e^{i\left( L\theta-\mu t/\hbar \right)}\{u_1(r)e^{i\left( l\theta-\omega t \right)}-v^*_1(r)e^{-i\left( l\theta-\omega^* t \right)}\}, \nonumber\\
    \label{eq:delta1} \\
    \delta\psi_2&=e^{i\left(-L\theta-\mu t/\hbar \right)}\{u_2(r)e^{i\left( l\theta-\omega t \right)}-v^*_2(r)e^{-i\left( l\theta-\omega^* t \right)}\}, \nonumber\\
    \label{eq:delta2}
  \end{align}
  where $l$ is the angular momentum quantum number.
  By inserting $\psi_j=\psi_j^0+\delta\psi_j$ to linearize the GP equations with respect to $\delta\psi _{j}$, we obtain the BdG equations. In matrix notation, these are
 \begin{eqnarray}
 \sigma {\cal M}{\it W}=\omega{\it W},
 \label{eq:bdg}
 \end{eqnarray}
where
 \begin{eqnarray}
 {\cal M}=\left(
 \begin{array}{cccc}
 h_{+} & Cn^0(r) & C_{12}n^0(r) & C_{12}n^0(r) \\
 Cn^0(r) & h_{-} & C_{12}n^0(r) & C_{12}n^0(r) \\
 C_{12}n^0(r) & C_{12}n^0(r) & h_{-} & Cn^0(r) \\
 C_{12}n^0(r) & C_{12}n^0(r) & Cn^0(r) & h_{+} \\
 \end{array}
 \right)
 ,\end{eqnarray}
  with
  \begin{align}
  h_{\pm}=&-\frac{1}{2}\left\{\frac{\partial^2}{\partial r^2}+\frac{1}{r}\frac{\partial}{\partial r}-\frac{(l\pm L)^2}{r^2}\right\} \nonumber \\
                 &+\frac{1}{2}r^2+2Cn^0(r)+C_{12}n^0(r)-\mu,
    \label{eq:h}
  \end{align}
  \begin{eqnarray}
    {\it W}=\left\{u_1(r),-v_1(r),u_2(r),-v_2(r)\right\}^T,
    \label{eq:w}
  \end{eqnarray}
  and $\sigma={\rm diag}(1,-1,1,-1)$.
  Here, the parameters are scaled as Eqs. (\ref{eq:scale}) and the tildes are omitted.
  Because the operator $\sigma {\cal M}$ in the BdG equations [Eq. (\ref{eq:bdg})] is non-Hermitian, the frequency $\omega$ may have an imaginary part.

  The linear stability of the system is investigated by numerically diagonalizing Eq. (\ref{eq:bdg}).
  The system is dynamically unstable when the frequency of excitations has an imaginary part ${\rm Im}\, \omega >0$ because the excitations are amplified exponentially with time.
  Here, we do not take into account the thermodynamic instability by neglecting energy dissipation, which causes spontaneous amplification of the collective modes with negative energy.
  Because a solution $(\omega,\ l,\ u_j, v_j)$ has its conjugate solution $(-\omega^*,\ -l,\ u_j^*, v_j^*)$, we present here only the results for $l \geqq 0$ without loss of generality.

 \begin{figure}[htb]
  \begin{center}
  \includegraphics[%
    width=1.0\linewidth,
    keepaspectratio]{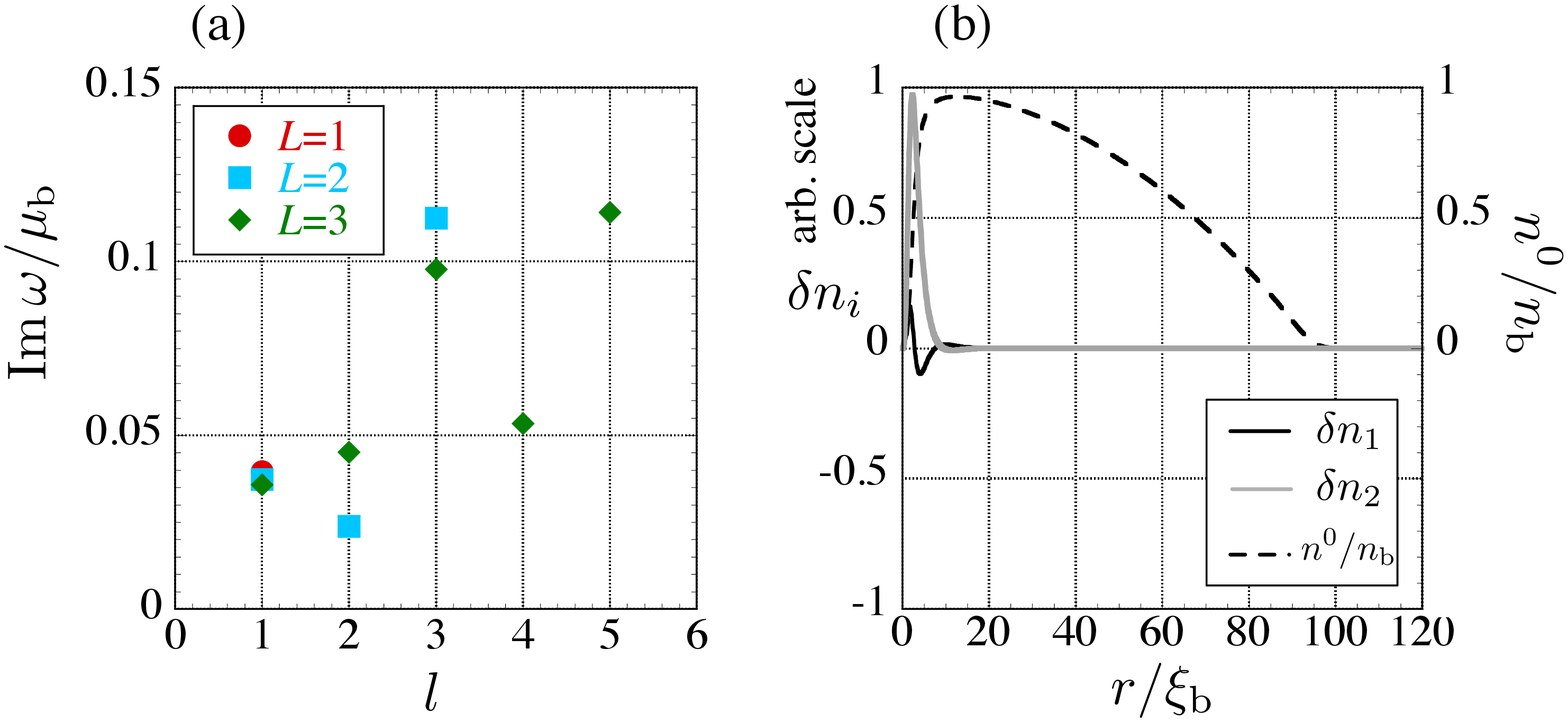}
  \end{center}
  \caption{(Color online)
    (a) Imaginary part of the frequencies of unstable excitations for $(1,-1)$-, $(2,-2)$-, and $(3,-3)$-vortices.
    (b) Radial distribution of the most unstable mode with $l=3$ for the $(2,-2)$-vortex.
    The solid line shows the change in density $\delta n_i$ caused by the most unstable mode.
    The dashed line shows the density profile $n_0$ in the stationary state.
   }
   \label{fig:bdg2}
 \end{figure}%


 \begin{figure}[htb]
  \begin{center}
  \includegraphics[%
    width=1.0\linewidth,
    keepaspectratio]{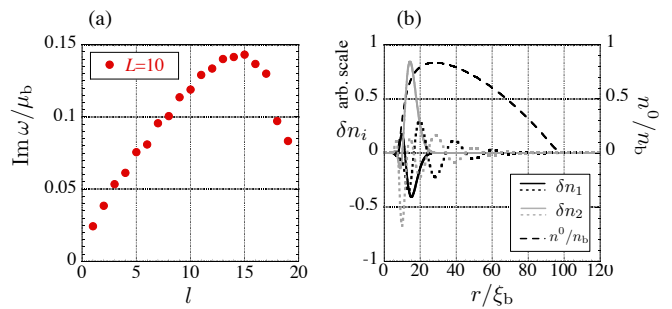}
  \end{center}
  \caption{(Color online)
    (a) Imaginary part of the frequencies of unstable excitations for $(10,-10)$-vortices.
    (b) Radial distribution of the most unstable mode with $l=15$ for the $(10,-10)$-vortex.
    The solid lines show the change in density $\delta n_i$ caused by the most unstable mode and the dotted lines show the radial profile of the change in density $\delta n_i$ caused by a characteristic unstable mode for the $(10,-10)$-vortex.
    The dashed line shows the density profile $n_0$ in the stationary state.
   }
   \label{fig:bdg10}
 \end{figure}%
 
  Figure \ref{fig:bdg2}(a) shows the imaginary part of the frequencies of unstable modes with ${\rm Im}\, \omega >0$ for the cases of small $L$: $(1,-1)$-, $(2,-2)$-, and $(3,-3)$-vortices.
  Although some unstable modes appear for each value of $l$, we show the largest imaginary part among them.

  First, we explain the case of $L=1$.
  This problem is connected to the interaction between vortices in different components in miscible two-component BECs.
  The interaction between vortices in different components is repulsive (attractive) for repulsive (attractive) interspecies interaction when the distance between the vortices is large compared to the size of the vortex core \cite{Eto_PRA2011}.
  Here, whether the intervortex interaction becomes repulsive or attractive is independent of the sign of the winding number of each vortex since the interaction results from the density nonuniformity through the term $g_{12}|\Psi_1|^2|\Psi_2|^2$ in the energy functional.
  However, our results show that the short-range interaction depends on the signs of the vortex winding numbers.

  $(1,-1)$-vortices have an unstable mode with $l=1$ [Fig. \ref{fig:bdg2}(a)].
  For an attractive ($g_{12}<0$) interspecies interaction, $(1,-1)$-vortices have an unstable mode with $l=1$, too.
  We found also that, for $(1,1)$- or $(-1,-1)$-vortices, there were unstable modes for $g_{12}>0$ but not for $g_{12}<0$.
  These results mean that the sign of the short-range interaction depends on the signs of the winding numbers of vortices for $g_{12}<0$; thus, the interaction between a $(1, 0)$-vortex and a $(0, 1)$-vortex is attractive but that between a $(1, 0)$-vortex and a $(0, -1)$-vortex is repulsive for attractive interspecies interaction.
  In fact, the amplitude of the unstable mode of a $(1,-1)$-vortex is localized around the vortex cores and its amplification makes the vortex split into a $(1,0)$-vortex and a $(0,-1)$-vortex.
  This effect is nontrivial compared to the long-range interaction between vortices in different components \cite{Eto_PRA2011}.

  Nontrivial effects also occur for $(L, -L)$-vortices with $L>1$.
  $(2,-2)$-vortices have unstable modes with $l=1,$ 2, and $3$.
  The mode with $l=3$ has the largest imaginary part and is thus the most unstable.
  Because the modes with $l$ cause the density profiles with  $l$-fold symmetry, it is expected that a density pattern with three-fold symmetry appears after onset of the instability.
  This situation  differs from the density patterns that appear in the splitting process of an $L$-charged vortex in single-component BECs, where an $L$-charged vortex splits into $L$ single-quantum vortices.
  Then the $L$-charged vortex has unstable modes with $l \leqq L$ that make a density pattern with $l$-fold symmetry.
  However, the number $l$ of the most unstable mode is larger than $L$ in the case of the CR vortices.

  Figure \ref{fig:bdg2}(b) shows the radial distribution of the most unstable mode for the $(2,-2)$-vortex, where we plotted the density fluctuation
  \begin{equation}
    \delta n_i(r) = |\psi_i^0(\theta=0,t=0)-\delta\psi_i^0(\theta=0,t=0)|^2 - n^0_i(r).
  \end{equation}
  The most unstable mode is localized in the vortex core and its amplitude decreases outside of the core.
  The amplitude vanishes at the center $r=0$ because of the divergence of the term $\propto (l\pm L)^2/r^2$ in Eq. (\ref{eq:h}) with $l\neq L$.
  The zero amplitude at $r=0$ makes it possible to cause vortices at $r=0$ after the amplification of the mode.
  In the case of the $(2,-2)$-vortex, the amplification makes a vortex with a winding number opposite to that of the original vortex in each component, as is discussed in Sec. \ref{sec:dynamics}.

  In Fig. \ref{fig:bdg2}(a), we plot the maximum values of the imaginary part ${\rm Im}\, \omega$ of the modes with $0 \leqq l \leqq 6$ for $L=1$, 2, and $3$ in BECs with a repulsive interspecies interaction.
  We also investigate the unstable modes for BECs with an attractive interspecies interaction.
  We observe that the imaginary parts of odd (even) $l$ are typically larger than those of even (odd) $l$ for $g_{12}>0$ ($g_{12}<0$).
  There is no physical explanation for this behavior at this time, and this is an open problem for the future.

  Next, we show a typical example of the instability of CR vortices with large $L$.
  $(10,-10)$-vortices have unstable modes with $l=1,2,\dots,19$ in Fig. \ref{fig:bdg10}(a).
  The most unstable mode has $l=15$.
  We show the change of the density caused by the most unstable mode (solid lines) and the typical unstable mode with $l=15$ for the cases of large $L$ (dotted lines) in Fig. \ref{fig:bdg10}(b).
  The amplitude of the most unstable mode is localized around the vortex core and decreases outside of the core as in the case of small $L$.
  However, the peak of the amplitude being almost outside of the vortex core differs from the case of small $L$.
  In addition to the localized modes, in this case, there appear unstable modes with large ${\rm Im}\,\omega$ whose amplitude is distributed broadly outside of the core.
  Typically, amplitudes of such modes oscillate spatially over a wide range.

  \subsection{Aspects of countersuperflow instability}\label{sec:csi}

  \begin{figure}[b!]
    \begin{center}
      \includegraphics[%
        width=1.0\linewidth,
        keepaspectratio]{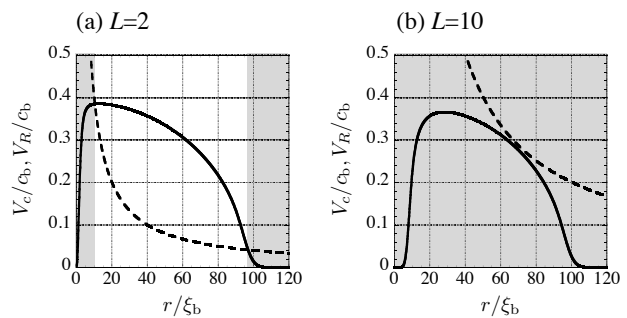}
    \end{center}
    \caption{
      (Color online)
      Radial profile of the local critical velocity $V_c$ (solid line) and the local relative velocity $V_R$ (dashed line) for the cases of (a) $L=2$ and (b) $L=10$.
      In the gray area, the relative velocity is larger than the critical velocity.
      The vertical axis shows the magnitude of the velocity normalized by $c_{\rm b}=\sqrt{\mu_b/m}$.
}
\label{fig:velocity}
\end{figure}

  We discuss here the relation between the instability of CR vortices and CSI.
  In Refs. \cite{Takeuchi_PRL2010,Ishino_PRA2011}, CSI has been discussed in the bulk where condensate densities are uniform.
  Characteristic aspects of CSI are expected to appear in our  CR vortex systems  because of the relative rotation between the two components.

  \begin{figure*}[t!]
    \begin{center}
      \includegraphics[%
      width=0.6\linewidth,
      keepaspectratio]{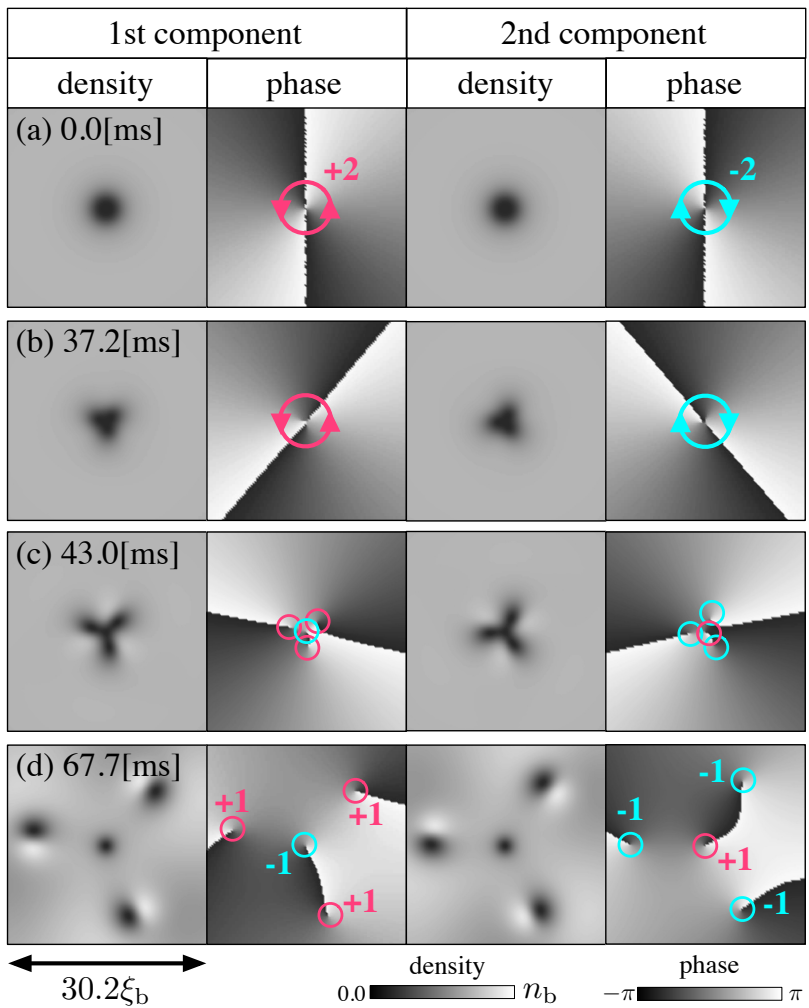}
    \end{center}
    \caption{
      (Color online)
      Time development of the vortex splitting of a $(2,-2)$-vortex
      (a) Initial state.
      (b), (c) A density pattern with three-fold symmetry appears owing to the strong amplification of the unstable mode with $l=3$.
      (d) Three $(1,0)$-vortices and three $(-1,0)$-vortices move away from the center.
           At the same time, a $(-1,1)$-vortex remains at $r=0$.
    }
    \label{fig:dy_el2-1}
  \end{figure*}
  
  To explain the nontrivial problem of the angular number $l$ of some unstable modes being larger than the winding number $L$ of the $(L,-L)$-vortex, we  further  proceed with the local density approximation.
  Let us introduce the local wave number vector ${\bm q}=(q_r, l/r)$ in polar coordinates, where $q_r$ is the pseudo-wave number in the radial direction.
  When the relative velocity $V_R$ is much larger than the critical velocity of CSI in uniform systems, the wave numbers $q_\parallel$ and $q_\bot$, which are, respectively, parallel and normal to the relative velocity, are characterized by the relation \cite{Ishino_PRA2011}
  \begin{align}
    (q_\parallel-V_R/2)^2+q_\bot^2=V_R^2/4.
  \end{align}

  As a first step of the analysis, we evaluate approximately the instability of CR vortices with the local density of the BECs.
  According to the form of the critical relative velocity of CSI in uniform systems,  we define the critical relative velocity $V_c$ in the local density approximation as
  \begin{align}
    V_{c}(r)=2\sqrt{Cn^0(r)(1-|\gamma|)},
    \label{eq:critical}
  \end{align}
  where $C$ is the nondimensional intraspecies interaction coefficient, $n^0(r)$ is the density profile of the stationary state, and $\gamma = C_{12}/C$.
  Equation (\ref{eq:critical}) and the local relative velocity $V_R(r)\equiv 2L/r$ are plotted for the $(2,-2)$-vortex and the $(10,-10)$-vortex in Fig. \ref{fig:velocity}.
  We have shown that an unstable mode has a certain amount of its amplitude even far from the vortex core for the $(10,-10)$-vortex.
  This must be interpreted in relation to CSI by the fact that the local relative velocity $V_R$ is larger than the critical velocity $V_c$ in the whole region for the $(10,-10)$-vortex [Fig. \ref{fig:velocity}(b)];
  CSI can occur locally in the bulk far from the $(10,-10)$-vortex because of the large relative velocity.
  In contrast, for the $(2,-2)$-vortex [Fig. \ref{fig:velocity}(a)], we have $V_R > V_c$ only near the vortex core and the surface of the BECs, where the local density approximation is inapplicable in the presence of a large gradient of the density.
  The fact that unstable modes are strongly localized in the vortex core in Figs. \ref{fig:bdg2}(b) and \ref{fig:bdg10}(b) is consistent with the large difference between $V_R$ and $V_c$ around the center $r \sim 0$, although we have found that the unstable modes do not appear near the surface.

  \begin{figure}[t!]
    \begin{center}
      \includegraphics[%
      width=0.8\linewidth,
      keepaspectratio]{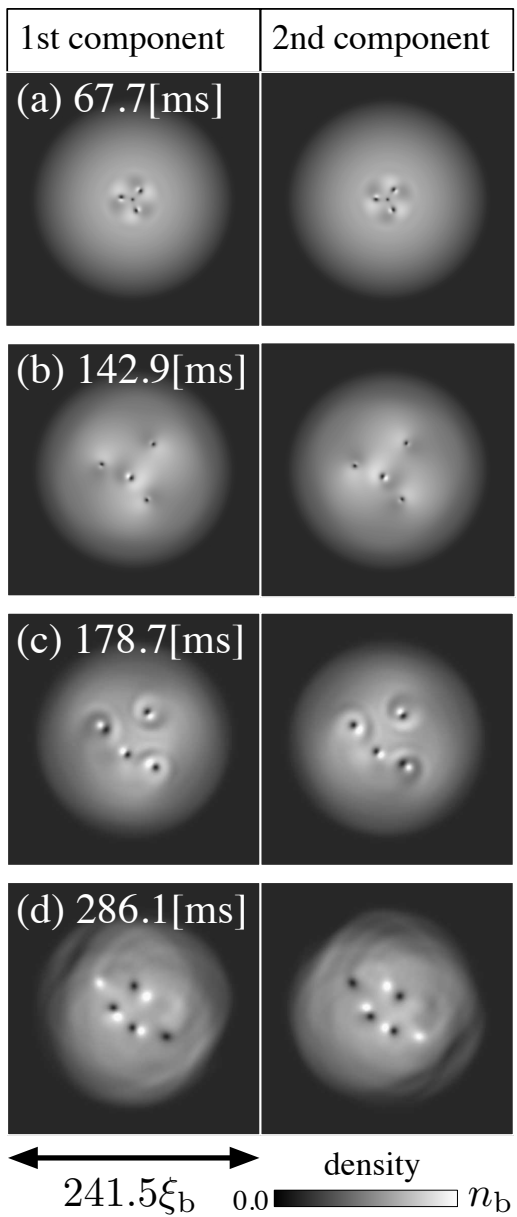}
    \end{center}
    \caption{
      Time development the late stage of the instability for the (2,-2)-vortex.
      The density profiles at $t=67.7 {\rm ms}$ (a), $t=142.9 {\rm ms}$, $178.7 {\rm ms}$, and $286.1 {\rm ms}$.
    }
    \label{fig:dy_el2-2}
  \end{figure}
  
  This argument can be applicable to our system qualitatively, because $V_R(r)$ is much larger than the criterion $V_c(r)$ in a broad area around the density peak for the $(10,-10)$-vortex, as shown in Fig. \ref{fig:velocity}(b).
  By replacing $q_\parallel$ and $q_\bot$ by $l/r$ and $q_r$, one obtains the characteristic numbers
  \begin{align}
    l &\sim r V_R=2L, \\
    q_r& \sim V_R/2=L/r.
  \end{align}
 The former number comes from the maximum value of $q_\parallel \sim V_R$ for unstable modes and the latter is the maximum wave number normal to the relative velocity.

  In fact, the former relation, $l=2L,$ is almost consistent with the maximum number $l=19$ of unstable modes for $L=10$ in Fig. \ref{fig:bdg10}(a).
  This relation also roughly describes the maximum   $l$ number even for the cases of small $L$ in Fig. \ref{fig:bdg2}(a).
  This consistency shows that the instability of CR vortices is dominated by CSI.
  In this way, the nontrivial unstable modes with $l>L$ obtained in the BdG model are qualitatively understood by CSI.

  Additionally, the radial wave number $q_r=L/r$ is roughly consistent with the wave number of the characteristic unstable modes for $(10,-10)$-vortices, which we show as dotted lines in Fig. \ref{fig:bdg10}(b).
  In an area around the density peak, $r\sim 30\xi_{\rm b}$, the radial wavelength is $\lambda_r = 2\pi/q_r\simeq 19\xi_{\rm b}$.
  This wavelength is consistent with the wavelength estimated from the characteristic unstable mode in Fig. \ref{fig:bdg10}(b).

  \section{Nonlinear development}\label{sec:dynamics}

  \begin{figure}[t!]
    \begin{center}
      \includegraphics[width=1.0\linewidth, keepaspectratio]{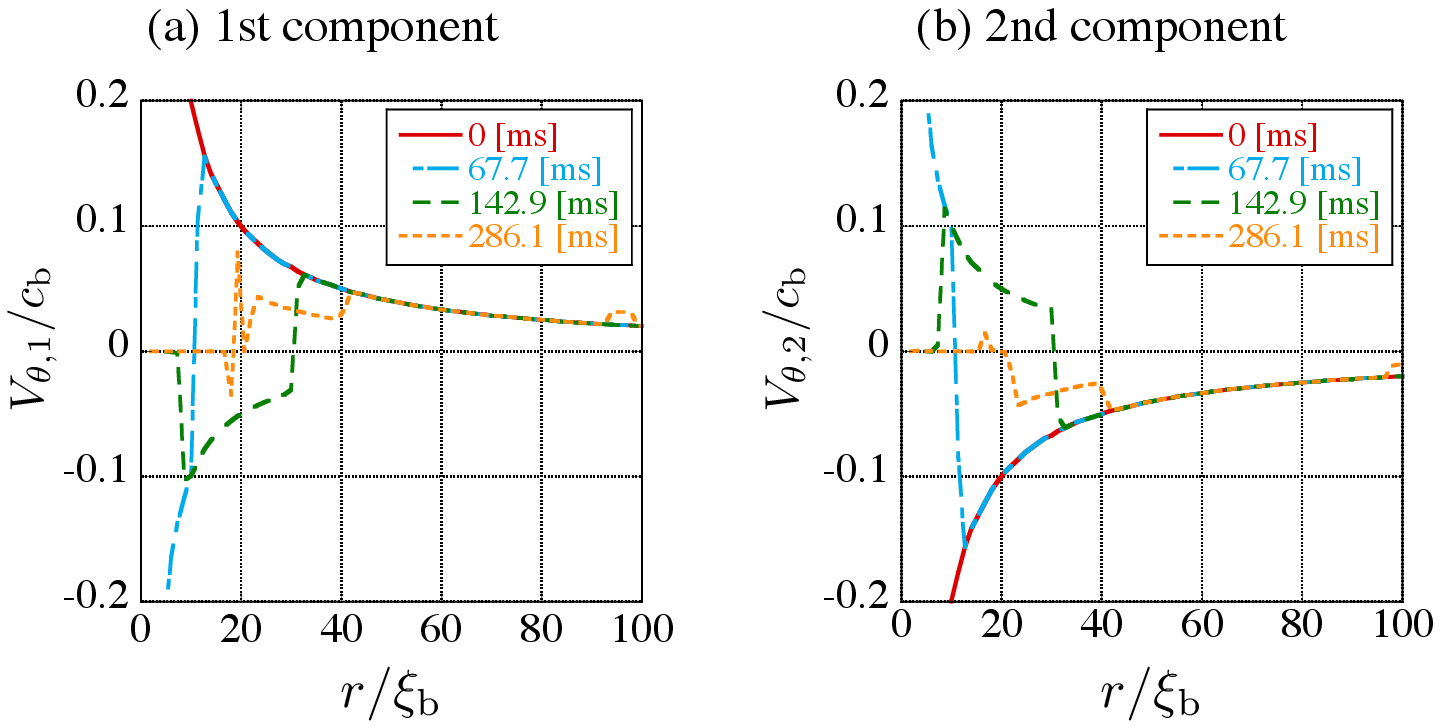}
    \end{center}
    \caption{
      (Color online)
      Time development of the averaged rotational velocity $V_{\theta,j}$ in the instability of a $(2,-2)$-vortex.
      The times $0 {\rm ms}$, $67.7 {\rm ms}$, $142.9 {\rm ms}$, and $286.1 {\rm ms}$ correspond to Figs. \ref{fig:dy_el2-1}(a), \ref{fig:dy_el2-1}(d), \ref{fig:dy_el2-2}(b), and \ref{fig:dy_el2-2}(d), respectively.
      The vertical axis shows the magnitude of the velocity normalized by $c_{\rm b}=\sqrt{\mu_b/m}$.
    }
    \label{fig:vr2}
  \end{figure}
  
  To reveal the nonlinear development of the instability of CR vortices, we numerically solved Eq. (\ref{eq:2DGP}).
  We consider a feasible case of binary BECs with repulsive interspecies interaction by using the same parameters as  in Sec. \ref{sec:formulations}.
  We investigated the time developments from the stationary states with small and large numbers of $L$ in Eqs. (\ref{eq:stationary1}) and (\ref{eq:stationary2}).
  To trigger the instability, a small white noise is added to the initial states.
  We do not demonstrate the time development of the instability from a $(1,-1)$-vortex, because the dynamics is simple; the amplification of the unstable mode with $l=1$ leads to the splitting of a $(1,-1)$-vortex into a $(1,0)$-vortex and a $(0,-1)$-vortex.
  We found that the splitting occurs even for $g_{12}<0$.
  Therefore, the short-range interaction between  $(1,0)$- and $(0,-1)$-vortices is considered to be repulsive for both $g_{12}>0$ and $g_{12}<0$.

  \begin{figure*}[t!]
    \begin{center}
      \includegraphics[width=1.0\linewidth, keepaspectratio]{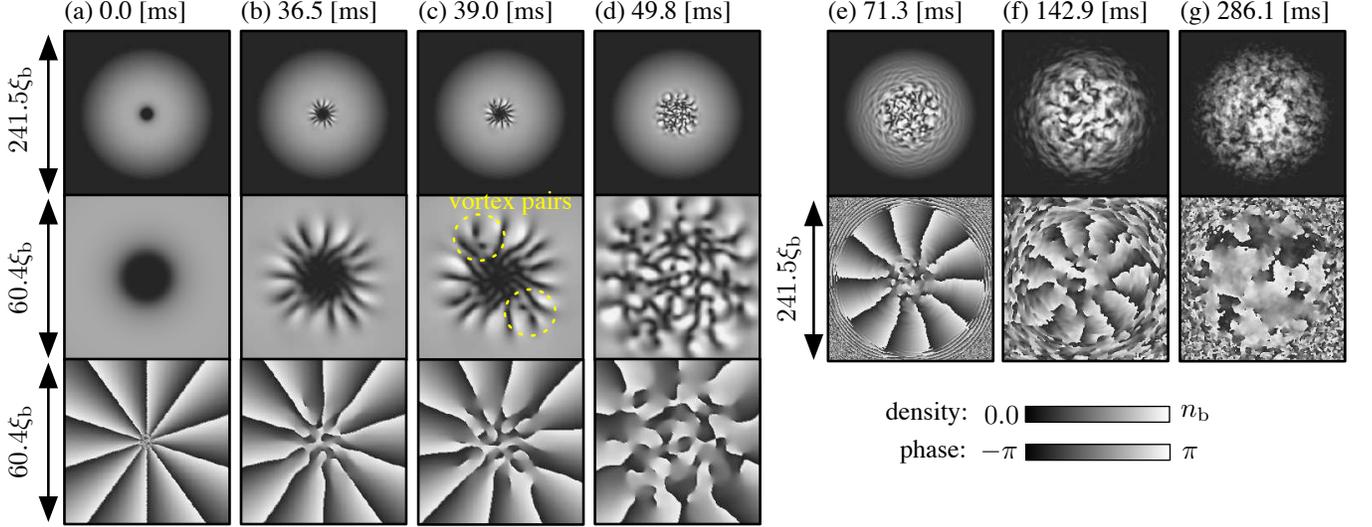}
    \end{center}
    \caption{
      (Color online)
      Nonlinear dynamics of the first component in the instability of a $(10,-10)$-vortex.
      The top panels show the density $|\psi_j({\bm r},t)|^2$.
      The middle and bottom panels of (a)--(d) show the close-up image of the density and phase around $r=0$.
      The bottom panels of (e)--(g) show the phase of the condensate.
      (a) Initial state.
      (b) A complex density pattern appears owing to the strong amplification of the unstable mode with $l=15.$
      The instability causes the characteristic density pattern with a $15$-fold symmetry around $r=0$.
      (c) Vortex pairs are nucleated in the regions far from $r=0$ so as to reduce the relative rotation.
      (d) A highly turbulent region appears around the center.
      (f) The turbulent region becomes larger.
      (g) The turbulent region spreads out to the whole system.
    }
    \label{fig:dy_el10}
  \end{figure*}

  \begin{figure}[htb!]
    \begin{center}
      \includegraphics[width=1.0\linewidth, keepaspectratio]{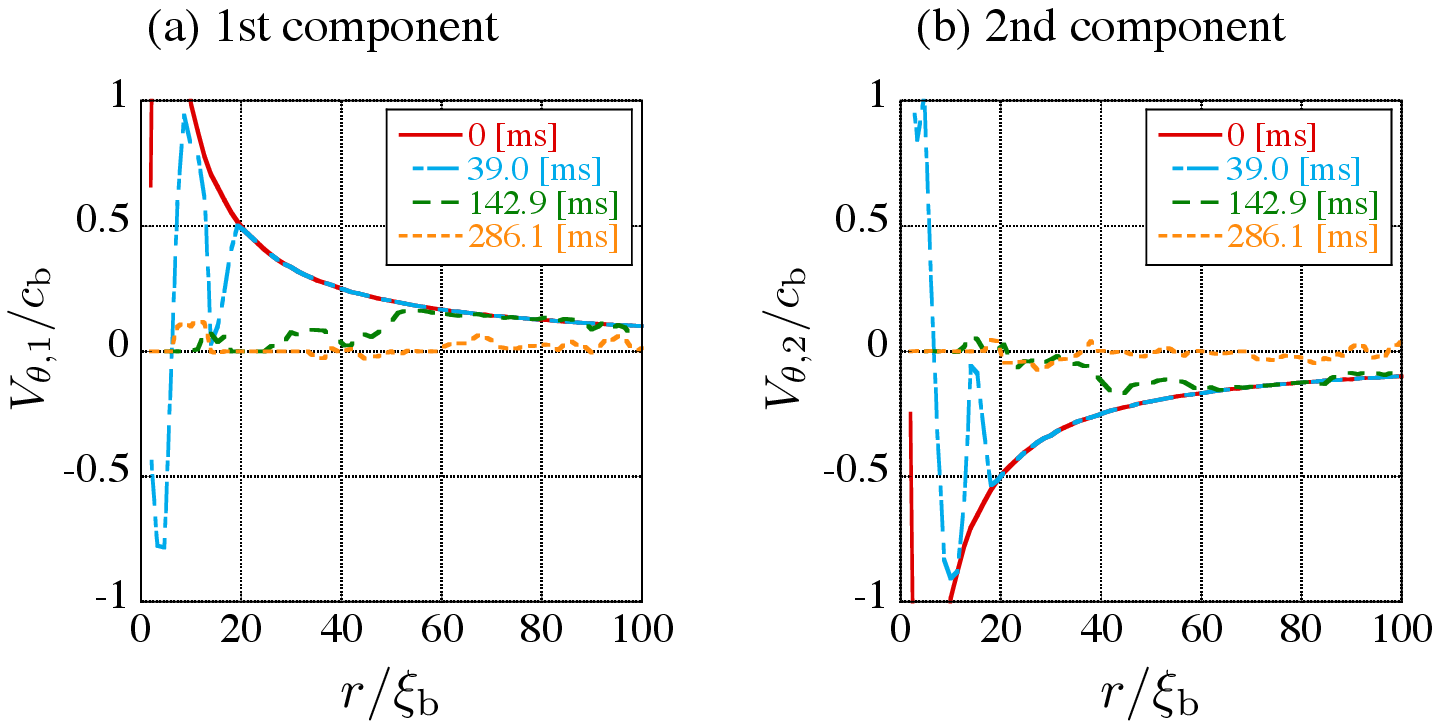}
    \end{center}
    \caption{
      (Color online)
      Time development of the averaged rotational velocity $V_{\theta,j}$ in the instability of a $(10,-10)$-vortex.
      The times $0 {\rm ms}$, $39.0 {\rm ms}$, $142.9 {\rm ms}$, and $286.1 {\rm ms}$ correspond to Figs. \ref{fig:dy_el10}(a), \ref{fig:dy_el10}(c), \ref{fig:dy_el10}(f), and \ref{fig:dy_el10}(g), respectively.
      The vertical axis shows the magnitude of the velocity normalized by $c_{\rm b}=\sqrt{\mu_b/m}$.
    }
    \label{fig:vr10}
  \end{figure}
  
  We will show first the instability dynamics of a $(2,-2)$-vortex as a typical example for the case of small $L$.
  Figure \ref{fig:dy_el2-1} represents the time development of the density and phase profiles of each component.
  In the early stage of the instability, a density pattern with three-fold symmetry appears owing to the strong amplification of the unstable mode with $l=3$ [Figs. \ref{fig:dy_el2-1}(a)--\ref{fig:dy_el2-1}(c)].
  Then, three single-quantum vortices move away from the center and a single-quantum vortex remains at $r=0$ in each component.
  The sign of the winding number of the vortex at $r=0$ is opposite to that of the three vortices.
  Thus, the total winding number is conserved throughout this process.
  Consequently, a $(2,-2)$-vortex splits into three $(1,0)$-vortices, three $(0,-1)$-vortices, and a $(-1,1)$-vortex.

  To understand the unique dynamics qualitatively, we calculated the quantity
  \begin{equation}
    V_{\theta,j}(r) = \left\langle{\bm v}_j\cdot{\bm e}_\theta \right\rangle_\theta,
  \end{equation}
   where ${\bm v}_{j} = (\psi_j^*\nabla\psi_j-\psi_j\nabla\psi_j^*)/2i|\psi_j|^2$ and ${\bm e}_\theta$ is the unit vector in the rotation direction.
  The brackets $\langle\cdots\rangle_\theta$ denote average over a circle of radius $r$.
  This quantity characterizes the radial profile of the mean local velocity in the rotational direction for the $j$th component.
  We have $V_{\theta,1}=L/r$ and $V_{\theta,2}=-L/r$ in the initial state and  $V_{\theta,1} \approx -V_{\theta,2}$ throughout the instability development because of the symmetric parameter setting between the two components.

  Figure \ref{fig:dy_el2-2} shows the time development after the process of Fig. \ref{fig:dy_el2-1}.
  The three pairs of  $(1,0)$- and $(0,-1)$-vortices move outward further [Figs. \ref{fig:dy_el2-2}(a) and  \ref{fig:dy_el2-2}(b)].
  Then, as shown in Fig. \ref{fig:vr2}, the relative rotational velocity between the two components is suppressed around the center, although its sign turns  negative there in the presence of a $(-1,1)$-vortex at $r=0$.
  The $(-1,1)$-vortex at the center is dynamically unstable, splitting into a $(-1,0)$-vortex and a $(0,1)$-vortex, both of which move outward [Figs. \ref{fig:dy_el2-2}(b) and  \ref{fig:dy_el2-2}(c)].
  After that, the relative velocity is suppressed and is almost zero in the center region (see Fig. \ref{fig:vr2}; $286.1 {\rm ms}$).
  We have observed that all vortices survive without pair annihilation until $286.1 \;{\rm ms}$ in the numerical simulation.

  The instability of CR vortices gradually becomes more complex when $L$ increases.
  The number of vortices that  appear after the vortex splitting process increases monotonically with $L$ according to the results of the linear stability analysis in Sec. \ref{sec:linear}.
  For example, a $(3,-3)$-vortex splits into seven vortices in each component, where we observed five $(1,0)$-vortices, five $(0,-1)$-vortices, two $(-1,0)$-vortices, and two $(0,1)$-vortices after the splitting process.

  If $L$ is large enough, the instability develops qualitatively different from that for small $L$.
  We have shown that some unstable modes can be distributed broadly far from the center $r=0$ for large $L$.
  These modes cause nucleation of vortices in the bulk region in addition to the vortex multiplication caused by vortex splitting in the center.
  In a three-dimensional homogeneous system, CSI causes nucleation of vortex rings after the characteristic density pattern formation \cite{Takeuchi_PRL2010,Ishino_PRA2011}.
  In our quasi-two-dimensional system, the instability causes pair nucleation of vortices in the bulk.

  Figure \ref{fig:dy_el10} shows the instability development from a $(10,-10)$-vortex.
  The most unstable mode in this case is $l=15$.
  The density pattern in the early stage [Fig. \ref{fig:dy_el10}(b)] is much more complex compared to that for $L=2$ in Fig. \ref{fig:dy_el2-1}.
  We can see in Fig. \ref{fig:dy_el10}(c) that vortex pairs are nucleated in the region far from $r=0$.
  Since the direction of superfluid velocity between a vortex and an antivortex of the vortex pairs is opposite to that of the initial rotational superflow in each component, the pair nucleation locally reduces the relative velocity $V_{\theta,1}-V_{\theta,2}$ around $r \sim 15 \xi_b$ in Fig. \ref{fig:vr10}.

  Because of the numerous vortices from pair nucleation in addition to  vortex splitting,  a highly turbulent region around the center appears  [Fig. \ref{fig:dy_el10}(d)].
  The relative velocity is strongly suppressed in the turbulent region and the region becomes larger with time [Figs. \ref{fig:dy_el10}(e) and \ref{fig:dy_el10}(f)].
  Eventually, the relative rotational velocity vanishes by the two components exchanging their angular momentum, and then the turbulent region spreads out to the whole system [Fig. \ref{fig:dy_el10}(g)].

  \section{Summary}

  We studied the linear stability and the instability development of CR vortices in miscible two-component BECs.
  We found that a CR vortex has unstable modes whose angular number is larger than the winding number of the CR vortex.
  The appearance of such modes is a unique feature of this system, which is dominated by CSI.
  The number of vortices  appearing in the vortex splitting process owing to the amplification of these modes is larger than the winding number of the initial vortex.
  The total winding number is conserved in this process by nucleating vortices with opposite winding number.
  When the winding number becomes larger, the unstable modes become more broadly distributed so as to nucleate vortex pairs in the bulk region.
  The vortices spread over the cloud, leading to binary quantum turbulence.
  The instability of CR vortices is one of the tools for creating binary quantum turbulence in BEC experiments.

  A CR vortex can be realized experimentally by applying the topological phase imprinting method \cite{Matthews_PRL1999,Williams_Nature1999,Shibayama_JPB2011}.
  We can imprint a phase that causes opposite rotations between two components by using two BECs with  different hyperfine states.
  Experimental evidence of the instability of CR vortices can be observed as the characteristic density pattern or the multiplication of vortices.
  Additionally, we  observe the drastic difference between the expansions of the cloud during the time of flight  before and after the instability, because the centrifugal force on the atoms is reduced by the relaxation of relative rotation caused by the instability.
  Experimental observation of the instability of CR vortices is valuable in terms of the physics of quantized vortices, hydrodynamic instability, and quantum turbulence.

  \begin{acknowledgements}
    S.I. acknowledges the support of a Grant-in-Aid for JSPS Fellows (Grant No. 244499).
    H.T. acknowledges the support of the ``Topological Quantum Phenomena'' (No. 22103003) Grant-in Aid for Scientific Research on Innovative Areas from the Ministry of Education, Culture, Sports, Science and Technology (MEXT) of Japan.
  \end{acknowledgements}



\begin{thebibliography}{99}

    \bibitem{Donnelly_Book1991} R. J. Donnelly, {\it Quantized Vortices in Helium I\hspace{-.1em}I } (Cambridge University Press, Cambridge, England, 1991).

    \bibitem{Vinen1957}W. F. Vinen, Proc. R. Soc. London, Ser. A {\bf 240}, 114 (1957); {\bf 240}, 128 (1957); {\bf 242}, 493 (1957); {\bf 243}, 400 (1958).

    \bibitem{Adachi_PRB2010} H. Adachi, S. Fujiyama, and M. Tsubota, Phys. Rev. B. {\bf 81}, 104511(2010).

    \bibitem{Kobayashi_PRL2005} M. Kobayashi and M. Tsubota, Phys. Rev. Lett. {\bf 94}, 065302 (2005).

    \bibitem{Kobayashi_PRA2007} M. Kobayashi and M. Tsubota, Phys. Rev. A {\bf 76}, 045603 (2007).

    \bibitem{Karn_PRB1980} P. W. Karn, D. R. Starks, and W. Zimmermann, Phys. Rev. B {\bf 21}, 1797 (1980).

    \bibitem{Matthews_PRL1999} M. R. Matthews, B. P. Anderson, P. C. Haljan, D. S. Hall, C E. Wieman, and E. A. Cornell, Phys. Rev. Lett. {\bf 83}, 2498 (1999).

    \bibitem{Shibayama_JPB2011} H. Shibayama, Y. Yasaku, and T. Kuwamoto, J. Phys. B: At. Mol. Opt. Phys. {\bf 44} 075302 (2011).

    \bibitem{Pethick_book} C. J. Pethick and H. Smith, {\it Bose-Einstein Condensation in Dilute Gases}, 2nd ed. (Cambridge University Press, Cambridge, England, 2008).

    \bibitem{Isoshima_PRL2007} T. Isoshima, M. Okano, H. Yasuda, K. Kasa, J. A. M. Huhtam\"{a}ki, M. Kumakura, and Y. Takahashi, Phys. Rev. Lett. {\bf 99} 200403 (2007).

    \bibitem{Shin_PRL2004} Y. Shin, M. Saba, M. Vengalattore, T. A. Pasquini, C. Sanner, A. E. Leanhardt, M. Prentiss, D. E. Pritchard, and W. Ketterle, Phys. Rev. Lett. {\bf 93} 160406 (2004).

    \bibitem{Mottonen_PRA2003} M. M\"{o}tt\"{o}nen, T. Mizushima, T. Isoshima, M. M. Salomaa, and K. Machida, Phys. Rev. A {\bf 68}, 023611 (2003).

    \bibitem{Kawaguchi_PRA2004} Y. Kawaguchi and T. Ohmi, Phys. Rev. A {\bf 70}, 043610 (2004).

    \bibitem{Huhtamaki_PRL2006} J. A. M. Huhtam\"{a}ki, M. M\"{o}tt\"{o}nen, T. Isoshima, V. Pietil\"{a}, and S. M. M. Virtanen, Phys. Rev. Lett. {\bf 97}, 110406 (2006).

    \bibitem{Skryabin_PRA2000} D. V. Skryabin, Phys. Rev. A {\bf 63}, 013602 (2000).

    \bibitem{Park_PRAR2004} Q.-H. Park and J. H. Eberly, Phys. Rev. A {\bf 70}, 021602(R) (2004).

    \bibitem{Brtka_PRA2010} M. Brtka, A. Gammal, and B. Malomed, Phys. Rev. A {\bf 82}, 053610 (2010).

    \bibitem{Wen_PRA2013} L. Wen, Y. Qiao, Y. Xu, and L. Mao, Phys. Rev. A {\bf 87}, 033604 (2013).

    \bibitem{Takeuchi_PRB2010} H. Takeuchi, N. Suzuki, K. Kasamatsu, H. Saito, and M. Tsubota, Phys. Rev. B {\bf 81}, 094517 (2010).

    \bibitem{Suzuki_PRA2010} N. Suzuki, H. Takeuchi, K. Kasamatsu, M. Tsubota, and H. Saito, Phys. Rev. A {\bf 82}, 063604 (2010).

    \bibitem{Sasaki_PRA2009} K. Sasaki, N. Suzuki, D. Akamatsu, and H. Saito, Phys. Rev. A {\bf 80}, 063611 (2009).

    \bibitem{Gautam_PRA2010} S. Gautam and D. Angom, Phys. Rev. A {\bf 81}, 053616 (2010).

    \bibitem{Takeuchi_PRL2010} H. Takeuchi, S. Ishino, and M. Tsubota, Phys. Rev. Lett. {\bf 105}, 205301 (2010).

    \bibitem{Ishino_PRA2011} S. Ishino, M. Tsubota, and H. Takeuchi, Phys. Rev. A {\bf 83}, 063602 (2011).
    
    \bibitem{Hamner_PRL2011} C. Hamner, J. J. Chang, P. Engels, and M. A. Hoefer, Phys. Rev. Lett. {\bf 106}, 065302 (2011).
    
    \bibitem{Abad_arXiv2013} M. Abad, A. Sartori, S. Finazzi, and A. Recati, arXiv:1310.0400.

    \bibitem{Eto_PRA2011} M. Eto, K. Kasamatsu, M. Nitta, H. Takeuchi, and M. Tsubota, Phys. Rev. A {\bf 83}, 063603 (2011).

    \bibitem{Williams_Nature1999} J. E. Williams and M. J. Holland, Nature (London) {\bf 401}, 568 (1999).

  \end{thebibliography}
\end{document}